\documentclass[twocolumn]{aastex63}
\shorttitle{Technosignature Drift Rate Distributions}
\shortauthors{Li et al.}
\graphicspath{{./}{figures/}}
\usepackage{amsmath}
\usepackage{wasysym}
\usepackage{mathrsfs}
\usepackage{hyperref}
\usepackage{verbatim}
\hypersetup{
    colorlinks=true,
    linkcolor=blue,
    filecolor=magenta,      
    urlcolor=blue,
    pdftitle={Overleaf Example},
    pdfpagemode=FullScreen,
    }
\begin{document}

\definecolor{cjg}{rgb}{0.0, 0.5, 0.5}
\definecolor{myh}{rgb}{0.5, 0.0, 0.5}
\newcommand{\cjg}[1]{\textcolor{cjg}{\bf [CJG: #1]}}
\newcommand{\myh}[1]{\textcolor{myh}{\bf #1}}

\title{Developing a Drift Rate Distribution for Technosignature Searches of Exoplanets}

\author[0000-0002-3012-4261]{Megan G. Li}
\affiliation{Department of Earth, Planetary, and Space Sciences, University of California, Los Angeles, CA 90095, USA}

\author[0000-0001-7057-4999]{Sofia Z. Sheikh}
\affiliation{SETI Institute, 339 Bernardo Ave, Suite 200 Mountain View, CA 94043, USA}
\affiliation{Berkeley SETI Research Center, University of California, Berkeley, CA 94720-3411}
\affiliation{Penn State Extraterrestrial Intelligence Center, 525 Davey Laboratory, The Pennsylvania State University, University Park, PA 16802, USA}

\author[0000-0002-1743-3684]{Christian Gilbertson}
\affiliation{Center for Computing Research, Sandia National Laboratories, Albuquerque NM 87185 USA}
\affiliation{Department of Astronomy \& Astrophysics, The Pennsylvania State University, 525 Davey Laboratory, University Park, PA 16802, USA}
\affiliation{Center for Exoplanets \& Habitable Worlds, The Pennsylvania State University, 525 Davey Laboratory, University Park, PA 16802, USA}

\author[0000-0002-5223-7945]{Matthias Y. He}
\affiliation{Department of Physics, University of Notre Dame, 225 Nieuwland Science Hall, Notre Dame, IN 46556, USA}
\affiliation{Department of Astronomy \& Astrophysics, The Pennsylvania State University, 525 Davey Laboratory, University Park, PA 16802, USA}
\affiliation{Center for Exoplanets \& Habitable Worlds, The Pennsylvania State University, 525 Davey Laboratory, University Park, PA 16802, USA}

\author[0000-0002-0531-1073]{Howard Isaacson}
\affiliation{Astronomy Department, University of California, Berkeley, CA}
\affiliation{Centre for Astrophysics, University of Southern Queensland, Toowoomba, QLD, Australia}

\author[0000-0003-4823-129X]{Steve Croft}
\affiliation{Berkeley SETI Research Center, University of California, Berkeley, CA 94720-3411}
\affiliation{SETI Institute, 339 Bernardo Ave, Suite 200 Mountain View, CA 94043, USA}

\author[0000-0001-5290-1001]{Evan L. Sneed}
\affiliation{Department of Earth and Planetary Sciences, University of California, Riverside, CA 92521, USA}
\affiliation{Berkeley SETI Research Center, University of California, Berkeley, CA 94720-3411}
\affiliation{Penn State Extraterrestrial Intelligence Center, 525 Davey Laboratory, The Pennsylvania State University, University Park, PA 16802, USA}

\begin{abstract}
A stable-frequency transmitter with relative radial acceleration to a receiver will show a change in received frequency over time, known as a ``drift rate''. For a transmission from an exoplanet, we must account for multiple components of drift rate: the exoplanet's orbit and rotation, the Earth's orbit and rotation, and other contributions. Understanding the drift rate distribution produced by exoplanets relative to Earth, can a) help us constrain the range of drift rates to check in a Search for Extraterrestrial Intelligence (SETI) project to detect radio technosignatures and b) help us decide validity of signals-of-interest, as we can compare drifting signals with expected drift rates from the target star. In this paper, we modeled the drift rate distribution for $\sim$5300 confirmed exoplanets, using parameters from the NASA Exoplanet Archive (NEA). We find that confirmed exoplanets have drift rates such that 99\% of them fall within the $\pm$53 nHz range. This implies a distribution-informed maximum drift rate $\sim$4 times lower than previous work. To mitigate the observational biases inherent in the NEA, we also simulated an exoplanet population built to reduce these biases. The results suggest that, for a Kepler-like target star without known exoplanets, $\pm$0.44 nHz would be sufficient to account for 99\% of signals. This reduction in recommended maximum drift rate is partially due to inclination effects and bias towards short orbital periods in the NEA. These narrowed drift rate maxima will increase the efficiency of searches and save significant computational effort in future radio technosignature searches.
\end{abstract}

\keywords{Search for extraterrestrial intelligence (2127), Exoplanets (498), Radio astronomy (1338)}

\section{Introduction}
\label{sec:intro}

The goal of astrobiology is to characterize the origins, evolution, and distribution of life in the universe \citep[e.g., ][]{desmarais2008astrobiology}. One particular strategy to search for life beyond Earth is to look for its ``technosignatures'', or astronomically-observable signs of non-human technology in the universe \citep{tarter2001search}. This strategy is often known as SETI, or the Search for Extraterrestrial Intelligence, and originated as an astrobiological sub-discipline in the 1960s, when it was realized that technological life in space could be detected in many ways: via its intentional electromagnetic communications \citep{cocconi1959communications,schwartz1961laser}, thermal waste from its megastructures \citep[][]{dyson1960spheres}, or even physical artifacts that it had sent to the solar system \citep[][]{bracewell1960probes}.

Modern developments in astronomical instrumentation --- for example, cutting-edge instruments such as TESS \citep{ricker2014transiting} and JWST \citep{gardner2006james} --- have inspired new technosignature search strategies \citep[e.g. ][for TESS and JWST respectively]{giles2021technosignatures,kopparapu2021nitrogen}. However, despite the proliferation of new search methods across the electromagnetic spectrum, most technosignature searches are still performed at radio frequencies \citep[e.g., ][]{price2020breakthrough}. This imbalance, while partially historical, is also well-motivated, as radio transmissions have many advantages that can be formalized with the Nine Axes of Merit from \citet{sheikh2020nine}. Radio technosignatures require no extrapolation from present-day human technology, often focus on signal morphologies with no astronomical confounders, and could contain large amounts of information transferred at the speed of light. In addition, searches for radio technosignatures are cheap to perform and can be easily executed in archival data.

``Narrowband'' signals are a particularly popular morphology in modern SETI searches, as they are efficient, used on Earth for communication, and have no natural confounders. To search for a narrowband signal, high spectral-resolution data are taken with a radio telescope, e.g. 3 Hz frequency resolution \citep[as in Breakthrough Listen's HSR data format and UCLA's searches, described by][respectively]{Lebofsky2019, margot2021search}. The time-frequency-intensity filterbank data product is often visualized via dynamic spectra or ``waterfall plots.'' Within those waterfall plots, a narrowband SETI signal would appear as a bright linear feature, a few channels wide, which drifts across frequencies over time --- this apparent slope is the ``drift rate'' of the signal. Drift rates are often, but not always\footnote{Drift rates can also be induced electronically, either deliberately or via uncorrected thermal or other effects in component electronics, as was the case in the signal-of-interest blc1 \citep[][]{Sheikh2021}.}, caused by relative acceleration between the receiver and the transmitter, implying that a zero-drift signal is likely radio frequency interference (RFI) in the same frame as the telescope. SETI researchers employ various algorithms to autonomously search for linear features with non-zero drift rates \citep[a summary of these methods can be found in ][]{sheikh2019choosing}; however, all of the popular algorithms require some limit to be placed on the absolute drift rate maximum, just as pulsar search algorithms require a maximum dispersion measure threshold \citep[e.g., ][]{parent2018implementation}. 

\citet{sheikh2019choosing} provided the first physically-motivated guideline for the choice of a maximum drift rate, based on contemporary knowledge of the orbital parameters of the most extreme exoplanetary systems. From this analysis, the authors found that a reasonable upper limit on the maximum drift rate, to account for all known exoplanets, was 200\,nHz, equivalent to a drift of 200\,Hz/s when observing at a frequency of 1\,GHz. This guideline, although comprehensive, was two orders-of-magnitude higher than the values being used in contemporaneous SETI searches, such as the $\pm$ 4 nHz searched by \citet{price2020breakthrough} or the $\pm$ 9 nHz searched by \citet{margot2021search}. Implementing a  200\,nHz maximum drift rate would require significantly more computational investment\footnote{Mitigated somewhat by the frequency-scrunching solution for drift rates past the ``one-to-one'' point described in \citet{sheikh2019choosing}}. 

In this work, we expand on \citet{sheikh2019choosing} by broadening the search to include \textit{all} points in an exoplanet's orbit (not just those that maximize the drift rate), to understand the distribution of expected drift rates in the galaxy below the previously-reported upper limit. We also sample a larger population of exoplanets discovered since the publication of \citet{sheikh2019choosing}, and begin to account for the drift rate effects of the biases present in the currently-known population of exoplanets. These distributions may be particularly useful to SETI searches without designated exoplanet targets, such as general star surveys and the galactic center. For searches focused on exoplanets themselves, it may be optimal to calculate the specific drift rate distributions for those exoplanets.

In Section \ref{sec:methods}, we will describe the method by which we calculate the drift rates and give a brief overview of the orbital parameters needed to calculate these drift rates. In Section \ref{sec:nea}, we will apply this method of calculating drift rates to exoplanets in the NASA Exoplanet Archive. In Section \ref{sec:fully_simulated}, we will apply the same methodology to a fully-simulated exoplanet population, in an attempt to mitigate some of the bias present in the observed exoplanet population. Finally, we will confirm the effects of the biases of the NEA sample on drift rates in 
\ref{sec:discussion}, and conclude in Section \ref{sec:conclusion}.
\section{Methods}
\label{sec:methods}

Imagine a situation where a radio receiver on Earth (at some latitude and longitude), points at a target star. That target star (with mass $M_{star}$) is host to an exoplanet (with some orbital parameters and physical properties). The exoplanet is host to a radio transmitter at some latitude and longitude on its surface.

In this model, a signal sent from the exoplanetary transmitter at $\nu_{rest}$ to the receiver on Earth will be received at some $\nu_{rec} = f(\nu_{rest}, t)$, where $f$ is some function constructed from the stellar, exoplanetary, and geometric parameters described above. The drift rate is the time-derivative of this function, $\dot{f}$.

Through this formulation, we can see that the total drift rate at any given moment is constructed from a combination of factors, including the orbit of the exoplanet around its host star, the rotation of the exoplanet hosting the transmitter, the movement of the exoplanet's host star relative to Earth, and the rotation and orbital motion of the Earth\footnote{Other contributions can be added if the transmitter is not fixed on the surface of an exoplanet, for example, the orbit of the transmitter around its host, the rotation of the transmitter about its own axis, and the inherent acceleration from propulsion on the transmitter itself \citep{sheikh2019choosing}. We neglect these terms here, as we assume a fixed surface transmitter.}.

Exoplanetary rotation rates are still relatively unknown and unconstrained.
\citet[][]{sheikh2019choosing} found that the orbital term is often much larger than the rotation term for exoplanets that are close to their host star. This is especially true for exoplanets orbiting M-dwarfs, which are typically expected to be tidally-locked. For this reason, we ignore the effects of rotation in this study.
Movement of the exoplanet's host star relative to Earth is also considered negligible \citep[][]{sheikh2019choosing}, so these drift rate contributions will also not be considered. In addition, the maximum contribution of the Earth's rotation (0.1\,nHz) and orbit (0.019\,nHz) around the Sun are known. We will not be taking these into account in this work, but they can be added back into the distribution we provide when doing a search.

This leaves us only with orbital contributions to drift rates from an exoplanet orbiting its host star. Drift rates are the result of Doppler acceleration: in an exoplanetary system where the mass of the host star is far greater than the mass of any of its exoplanets, the only acceleration we need to consider is the gravitational force of the host star exerted onto the exoplanet. Then, the drift rate can be expressed as:

\begin{equation}
    \dot{\nu} = \frac{\nu_{rest}}{c} \frac{GM_{star}}{r^2} sin(i)
\end{equation}
where $i$ is the inclination of the orbit relative to the sky plane and $r$ is the instantaneous distance between the exoplanet and its host star. Remember that $\nu_{rest}$ is the frequency that is being transmitted, as measured by the transmitter. Given that modern ultra-wideband receivers \citep[such as that described by][]{price2021expanded} cover enough bandwidth that the drift rate limit, defined this way, would change by a factor of a few across the band, we will generalize this equation so that is independent of rest frequency:

\begin{equation}
    \dot{\nu}_{norm} = \frac{\dot{\nu}}{\nu_{rest}}= \frac{GM_{star}}{cr^2}sin(i)
\end{equation}

$\dot{\nu}_{norm}$ is a normalized drift rate, with units of Hz, which we will refer to, for simplicity, as just the ``drift rate'' for the rest of this paper.

\subsection{Calculating Drift Rates Given Orbital Parameters}
\label{ssec:the_code}

For a full set of exoplanetary orbital parameters --- semimajor axis, period, inclination, eccentricity, argument of periastron, and longitude of ascending node --- we can calculate the drift rate as seen from Earth at any point in the exoplanet's orbit. 

This orbit can be modeled mathematically using gravitational force and conservation of angular momentum. The following derivation is an approximation for a single planet system. We begin with Newton's Law of Universal Gravitation, 
\begin{equation} \label{eq:newton}
    F(r) = G \dfrac{m_1 m_2}{r^2}
\end{equation}

where $m_1$ and $m_2$ are the masses of the star and the exoplanet, and $r$ is the distance between them.

To apply the conservation of angular momentum, we can use Lagrangian mechanics, 
\begin{equation}
    \mathcal{L} = \dfrac{m_1 m_2}{2(m_1 + m_2)}(\dot{r}^2 + r^2 \dot{\phi}^2) - U(r)
\end{equation}
and 
\begin{equation}\label{eq:derivatives}
    \dfrac{d\mathcal{L}}{dr} = \dfrac{d}{dt} \dfrac{d\mathcal{L}}{dr}
\end{equation}

Substituting for the derivatives in equation \ref{eq:derivatives},
\begin{equation} \label{eq:solved_lagrange}
    \dfrac{m_1 m_2}{2(m_1 + m_2)}r\dot{\phi}^2 -\dfrac{dU}{dr} = \dfrac{m_1 m_2}{2(m_1 + m_2)}\ddot{r}
\end{equation}
Thus,
solving for $r$ from equations \ref{eq:newton} and  \ref{eq:solved_lagrange} to obtain a function for the distance between the exoplanet and the star as a function of angular position $(\phi)$, we obtain the general solution that 
\begin{equation}
r(\phi) = \dfrac{a (1-e^2)}{1+ecos\phi}
\end{equation}
where a is the semimajor axis and $e$ is the eccentricity.

\begin{figure*}
    \centering
    \includegraphics[scale = 0.4]{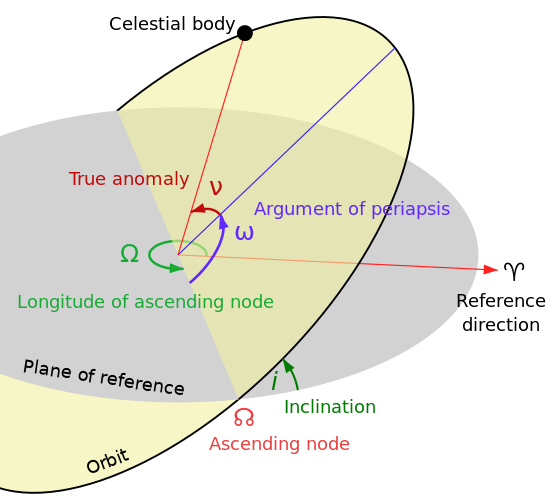}
    \caption{This figure shows an example orbit of an exoplanet. The reference direction, $\aries$, is pointed toward Earth. Periapsis refers to the point of the orbit in which the orbiting body is closest to the body it orbits. For our purposes, this is always the periastron, as our exoplanets are orbiting stars. Figure from \href{https://commons.wikimedia.org/wiki/File:Orbit1.svg}{Wikimedia Commons.}}
    \label{fig:orbit}
\end{figure*}

This orbit is displayed in Figure \ref{fig:orbit}, where $\aries$, the reference direction, points toward Earth. 
The angle between $\aries$ and the plane of the orbit is $i$, the inclination. 
The ascending node is the point where the exoplanet rises above the plane of reference.
Thus, the longitude of the ascending node, $\Omega$, is the angular position of the ascending node on our plane of reference, where $\aries$ points from where $\Omega = 0$. 
Finally, the argument of periastron, $\omega$, is the angle between the ascending node and the orbit's periastron.

We generated the elliptical orbit of each exoplanet as described above and sampled the gravitational acceleration $\frac{GM}{r^2}$ at 200 evenly spaced times throughout one  orbital period. Note that evenly sampling in time, instead of in angular position, results in more samples being taken near apoastron as opposed to periastron. We then project these accelerations in the direction of the observer to obtain the normalized drift rate. 


\section{Drift rates from the NASA Exoplanet Archive}
\label{sec:nea}

\subsection{Missing Orbital Parameters}
\label{ssec:nea_missing_paramters}

The NASA Exoplanet Archive (NEA) \citep[][]{ps} is an online database providing information on confirmed exoplanets and exoplanet candidates \citep{NEA2013}. We chose to use the NEA sample because it is the most comprehensive publicly-available archive for exoplanets across multiple missions. We use the archive as it was in 2023 May, when it contained 5347 exoplanets. Unfortunately, the exoplanet search methods and missions that supply much of the NEA sample do not (and cannot) solve for all seven orbital parameters for the exoplanets that they detect, and are particularly insensitive to the argument of periastron and the longitude of ascending node. We describe below how we account for missing values.

\subsubsection{Period or Semimajor Axis}
\label{sssec:nea_kepler_drawing}

Kepler's Third Law gives us a way to derive a planet's semi-major axis $a$ from its period $P$, or vice versa, so long as we know the host star's stellar mass $M_{star}$. We can insert values from the NEA sample into $P^2 = \frac{4\pi^2a^3}{G(M_{star})}$ assuming $M_{star} >> M_{planet}$. Out of 5347 planets in the NEA sample, 61 are removed because only one of [$P$, $a$, $M_{star}$] are present, leaving 5286 exoplanets in the sample. 2192 exoplanets had no semimajor axis recorded, 236 had no period recorded, and 748 had no stellar mass recorded.

\subsubsection{Inclinations}

3899 exoplanets in the remaining 5286 have missing inclinations. However, the discovery method for all of these exoplanets is available, so for any exoplanet discovered through either the transit or radial velocity method, we set $i = 90^\circ$, as both of these methods are most sensitive to $i \sim 90^\circ$. Any of the few remaining exoplanets that needed to be assigned an inclination were given a random selection from $-1 \leq cos(i) \leq 1$.

\subsubsection{Arguments of Periastron and Longitude of Ascending Node}
\label{sssec:nea_distribution_drawing}
4064 of the 5286 exoplanets were missing arguments of periastron, and fewer than 10 have longitudes of ascending node. Neither of these two parameters have preferred values, as they are geometrical (not physical) parameters measured with respect to Earth. Therefore, when one or both of these parameters is
missing, we use a randomly-selected number from a uniform distribution of $0^\circ - 360^\circ$. 

\subsubsection{Eccentricity from Rayleigh Distribution}
For exoplanets that did not have a recorded eccentricity, we randomly selected eccentricities from a Rayleigh distribution with $\sigma=0.02$, according to estimates for multiplanet systems from \citet[][]{he2019architectures}; transit surveys suggest that most exoplanets reside in multi-planet systems \citep[e.g., ][]{batalha2013planetary, mulders2018exoplanet}. 3256 exoplanets had no recorded eccentricity. The Rayleigh distribution from which the missing eccentricities are drawn is described by:
\begin{equation}
f(e) = \dfrac{e}{0.004}exp(-\dfrac{e^2}{0.008}) \end{equation}
The mean and variance of this distribution are 0.025 and 0.00017, respectively.

\subsection{Drift Rate Histogram from the NEA Sample}

\begin{figure*}[ht]
    \centering
    \includegraphics[scale = 0.6]{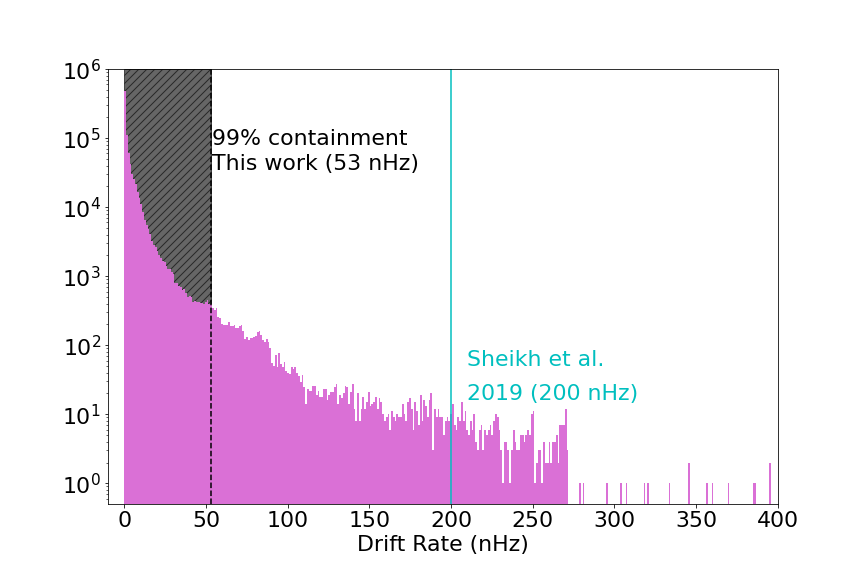} 
    \caption{A histogram of all of the calculated drift rate magnitudes for the 5286 planet sample from the NASA Exoplanet Archive, with drift rates calculated for each planet at 200 evenly-spaced times in the orbit. This histogram therefore contains 1,057,200 drift rates (200 for each of the 5286 exoplanets). The dotted line at $53$ nHz designates the 99\% boundary: 99\% of calculated drift rates fall between 0 and this absolute value. The solid line demonstrates the suggested drift rate upper limits from \citet[][]{sheikh2019choosing}. Note that for 99\% of the known exoplanet sample, the \citeauthor{sheikh2019choosing} drift rate is at least 4 times larger than necessary.}
    \label{fig:drift_rate_histogram}
\end{figure*}

As described in Section \ref{sec:methods}, we used the orbital parameters acquired or generated above to produce 200 drift rates, equally-spaced in time for each orbit, for each of the 5347 planets with known parameters in the NEA sample. We compiled these drift rates into a histogram (Figure \ref{fig:drift_rate_histogram}). Figure \ref{fig:drift_rate_histogram} illustrates that a narrowband radio technosignature search with a maximum drift rate of $\pm 53$\,nHz would catch 99\% of the drift rates produced by known exoplanets. In contrast, the upper-limit suggested by \citet[][]{sheikh2019choosing}, at $\pm 200$\,nHz, captures $<1\%$ more drift rates for four times the computational cost\footnote{Assuming a naive linear scaling; see \citet{sheikh2019choosing} for a more detailed treatment of the issue.}, which is likely not an effective trade-off for most applications. This factor of four difference between our suggested range and the range of \citet[][]{sheikh2019choosing} is primarily due the distribution of inclinations (\citeauthor{sheikh2019choosing} choose $90\degr$ for all bodies) and of semimajor axes (\citeauthor{sheikh2019choosing} choose the smallest possible semimajor axis). 

\section{Drift rates from a fully-simulated exoplanet population}
\label{sec:fully_simulated}

\subsection{The NEA sample is not fully representative of the underlying exoplanet population}

As of May 2023, most exoplanets in the NASA Exoplanet Archive (NEA) sample have been discovered through either the transit method (75\%), in which a planet periodically dims the host star's light as it passes in between the observer and its host star, or the radial velocity (RV) method (20\%), in which the slight wobble of a star due to its planet's gravitational pull is measured via shifting spectral features\footnote{\url{https://exoplanetarchive.ipac.caltech.edu/docs/counts_detail.html}}. Both of these methods have detection biases which affect their completeness across parameter space, which have been well-documented in the literature \citep[e.g.,][]{christiansen2020measuring, teske2021magellan}.

The transit method inherently detects a biased sample of exoplanets. For example, systems that have larger planets in comparison to their host star display a strong luminosity difference should a planet transit, making them easier to detect. Duration of transit and the number of transits available in an observation window also bias the population. In addition, the transit method is only sensitive to inclinations near $90^\circ$, as the planet must transit between Earth and the host star to be detected. The RV method does not require such restrictive inclinations, but is also more sensitive to near edge-on configurations which maximize the RV semi-amplitude. However, it involves a whole host of additional detection biases related to target selection, data coverage, and model fitting. To assess how much these biases affect the NEA sample, and how that might inform the \textit{overall} drift rate distribution of the underlying exoplanet population (not just that of known exoplanets), we constructed a simulated exoplanet population which corrects for some of these biases.

\subsection{Distributions for Simulated Parameters}
\label{ssec:sim_parameters}

To simulate an exoplanet population, we need to construct un-biased distributions for the orbital parameters described in Section \ref{sec:methods} as well as stellar mass, and then sample from those distributions. Below, we describe the chosen distributions for each parameter.

\subsubsection{Inclination}
Perhaps the most notable difference between the NEA sample and the underlying exoplanet population is that their inclinations should be uniformly distributed such that $cos(i)$ is between $-1$ and $1$, as described in Section \ref{sssec:nea_distribution_drawing}. We used this distribution in the simulation.

\subsubsection{Longitude of Ascending Node and Argument of Periastron} These parameters were drawn from a uniform distribution over $0^\circ$ to $360^\circ$ --- this is the same treatment we applied in Section \ref{sssec:nea_distribution_drawing} when these parameters were missing in the NEA database. 

\subsubsection{Stellar Mass}

We drew stellar masses randomly from the exoplanetary hosts in the Gaia–Kepler Stellar Properties Catalog \citep{berger2020gaia}, which includes 186,301 Kepler stars. Kepler searched F, G, K, and M type stars with an emphasis on F and G types \citep{berger2020gaia}.

\subsubsection{Period and Semimajor Axis} We drew periods from a broken power law of $P^{-1.5}$ from 5 hours to 10 days and $P^{-0.3}$ from 10 days to 300 days, as informed by models from \citet[][]{he2019architectures}, \citet{winn2018kepler} and \citet{mulders2018exoplanet}. We then calculated semimajor axes from each individual planet's host star mass and period using Kepler's Third Law. The model described by \citet{he2019architectures} was trained on F, G, and K type stars, which is congruent to our stellar mass distribution from \citet{berger2020gaia}.

\subsubsection{Eccentricity} It is currently understood that the underlying eccentricity distribution of exoplanets is dependent on whether they reside in a single-planet system or a multi-planet system, with single-planet systems having larger eccentricities on average \citep[][]{he2019architectures}. While we used these two eccentricity distributions, our simulations did not account for planet-to-planet interaction. We will refer to the populations drawn from these two distributions as the low eccentricity population and the high eccentricity population. We drew the low eccentricity population from a Rayleigh distribution with $\sigma = 0.02$ (as described in Section \ref{sssec:nea_distribution_drawing}). The high eccentricity population can also be described with a Rayleigh distribution, but with $\sigma = 0.32$ \citep[][]{he2019architectures}. We ran two sets of simulations, calculating drift rates for both of these eccentricity models, to evaluate the effect of the eccentricity distribution on the drift rate results.

\subsection{Drift rate Histograms from the Fully-Simulated Population}

\begin{figure*}
    \centering
    \includegraphics[width=\textwidth]{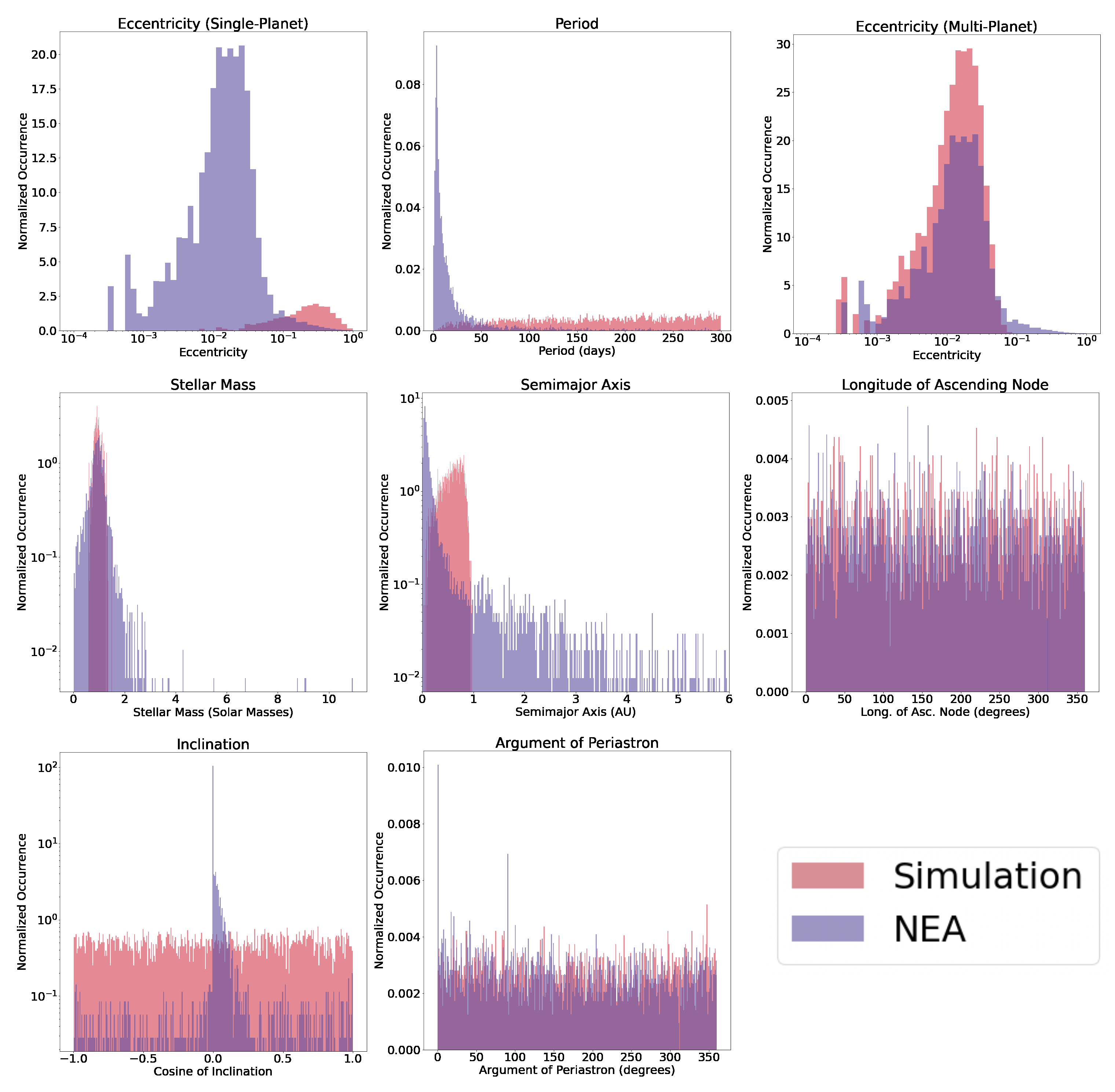}
    \caption{Histograms comparing the inputs for each of the 7 parameters used to calculate drift rates, with low-eccentricity and high-eccentricity options being shown in separate subplots. The simulated parameters are in red, and the NEA sample's parameters are in blue. The x axes for period and semimajor axes are only shown up to 300 days and 6\,AU, respectively, for visualization purposes, as mathematical models currently only predict this far. However, the NEA sample has entries past these limits.}
    \label{fig:distribution_plots}
\end{figure*}

Using the distributions described in Section \ref{ssec:sim_parameters}, we created twenty simulated populations of 5286 exoplanets each: ten with low eccentricities and ten with high eccentricities. For each of the ten simulations of each population, there was nominal variance. As an  illustration, in Figure~\ref{fig:distribution_plots}, we show the parameter distributions from a single simulation with each eccentricity distribution compared to the NEA sample. 

Each of the parameters was drawn randomly from its associated distribution, and the draws were performed independently. There is some dependence between planetary parameters, especially for planets in multi-planet systems (due to, e.g., orbital stability, the so-called `peas-in-a-pod' patterns describing the correlated sizes and spacings of planets in the same system \citep{he2020architectures,https://doi.org/10.48550/arxiv.2203.10076}. However, because the interplay between these parameters is an active and complex area of study \citep[see, e.g., ][]{he2020architectures}, a detailed treatment is outside of the scope of this work.

\begin{figure*}
    \centering
    \includegraphics[width =\textwidth]{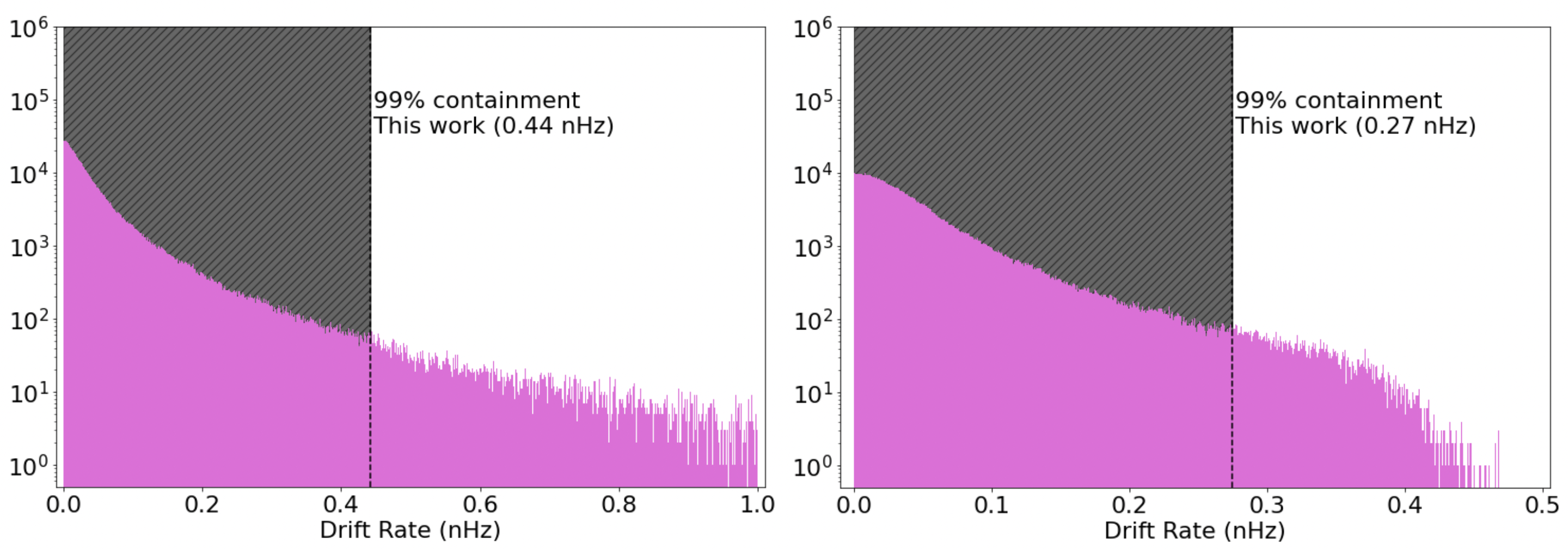}
    \caption{Histograms for the entirely simulated populations of exoplanets (high eccentricities on the left, low eccentricities on the right). Each population has 5286 planets with 200 drift rates drawn for each planet, giving a total of 1,057,200 drift rates, to mirror Figure \ref{fig:drift_rate_histogram}. The dotted lines on each histogram give the 99\% containment boundary. For high eccentricity, 99\% of the drift rates fall in $\pm 0.44$\,nHz. For low eccentricity, 99\% of the drift rates fall in $\pm 0.27$\,nHz. }
    \label{fig:sim_hists}
\end{figure*}

We computed drift rates at 200 points in time in each planet's orbit for each of the 5286 simulated planets and once again visualized the drift rate distributions via histograms. The histograms of drift rates for high eccentricity populations and low eccentricity populations are shown in Figure \ref{fig:sim_hists}. Overall, calculating drift rates for the simulated, ``debiased'' exoplanet distributions results in maximum drift rate thresholds (at the $99^{th}$ percentile) which are \textit{over two orders of magnitude} lower than the results from the NEA sample, illustrating the overwhelming effect of observational bias.

\section{Discussion}
\label{sec:discussion}

\subsection{Assessing differences between NEA sample and simulated samples}
\label{ssec:comb_sim_with_90}

In Section \ref{sec:fully_simulated} we discovered that a fully-simulated exoplanet population, with observational biases accounted for, produces a drift maximum at the $99^{th}$ percentile which is more than two orders-of-magnitude smaller than that of the NEA sample. These biases are visualized in Figure \ref{fig:distribution_plots}. Histograms of how each bias affects the distribution of drift rates are available in the Appendix.

The bias in inclination is due to discovery method as described in Section \ref{sec:fully_simulated}. De-biasing the inclination alone, while keeping the rest of the NEA sample parameters the same as in Section \ref{sec:nea}, reduced our 53~nHz for 99\% threshold to 42~nHz. It is believed that most exoplanets are in multi-planet systems, which accounts for the NEA sample's low eccentricities \citep[e.g., ][]{batalha2013planetary, mulders2018exoplanet}. Changing the NEA eccentricity distribution to the simulated low-eccentricity population resulted in a negligible change in drift rate threshold. 

The difference in period and semimajor axis is two-fold: our period model has an upper limit of 300 days, while the NEA has no such restriction, and the NEA sample is biased toward planets with short periods (as exoplanets with shorter periods are easier to detect with both transits and RVs). Both effects can be attributed to the requirement that more than one orbital period of observation is necessary to confirm the existence of an exoplanet. This difference, particularly the bias towards short-period planets in the NEA sample, appears to have had the strongest effect on our results.

\subsection{Outliers and maximum drift rates}

One of the  simulations with a single-planet eccentricity distribution produced a drift rate of $10^4$\,nHz, the largest value that appeared in any of our calculations. This is a good reminder that a transmitter on a planet with a large eccentricity, close to its host star, and seen at exactly the wrong angle and time with respect to an observer on Earth could produce a drift rate large enough to exceed any maximum drift rate guideline that is not a physical upper limit \citep[as discussed by][]{sheikh2019choosing}. However, when a balance must be struck between parameter space covered and logistical considerations, using a drift rate maximum for which an algorithm could find a direct drift rate match in 99\% of scenarios is entirely sufficient.

\subsection{Drift rate equivalents in the optical}

Technosignature searches in the optical and infrared are not limited by the availability of compute power to the same degree as radio SETI searches. Where radio searches comb through time-resolved data products, optical SETI searches utilize either time-compressed data, such as high-resolution spectra \citep{Zuckerman2023} or nano-pulses that do not have resolved energies \citep{Foote2021,Wright2019}. Both types of search allow for complete examination of parameter space without being computation limited.  

When identifying the position of a candidate laser emission line in high-resolution spectra, the radial velocity between the emitter and detector will determine the blueshift or redshift of the observed laser. Considering the Earth alone, the largest contribution to the radial velocity is the revolution of the Earth around the sun, at 30\,km\,s$^{-1}$. If a laser emission line was detected while the Earth is at opposite sides of its orbit, and in the ecliptic, the resulting difference would be 60\,km\,s$^{-1}$. For an optical spectrometer with a resolving power of $50,000-100,000$, each pixel represents $1-2$\,km\,s$^{-1}$. The candidate laser line would move many tens of pixels due to the Earth’s motion, independent of the motions of the emitter. Since we understand the barycentric motions of the Earth, we can subtract off the Earth’s contribution to determine the emitter’s radial motion relative to us.

\section{Conclusion} 
\label{sec:conclusion}

In Section \ref{sec:intro}, we discussed the motivations and background behind quantifying drift rates. Then, in Section \ref{sec:methods}, we described the theory and methodology behind quantifying drift rates. In Section \ref{sec:nea}, we calculated drift rates for known exoplanets in the NASA Exoplanet Archive. To address some of the biases in the NEA, we then simulated a population of exoplanets and calculated their drift rates in Section \ref{sec:fully_simulated}. Section \ref{sec:discussion} discusses some of the implications of this work and reconciles the differences between the drift rates of the NEA sample and those of the simulation. 

\citet{sheikh2019choosing} suggested limits of $\pm 200$\,nHz for radio technosignature searches.  Exoplanet parameters from the NASA Exoplanet Archive yield 99\% of drift rates in the range $\pm 53$\,nHz. A simulated population of exoplanets motivated by population models \citep{mulders2018exoplanet, he2019architectures} yields 99\% of drift rates between $\pm 0.44$ (0.27) nHz assuming high (low) eccentricities. These simulations indicate that a more complete survey of exoplanets that encompasses a wider distribution of inclinations would significantly lower the drift rate distribution of known exoplanets to the $\pm 0.5$\,nHz range.

The implications of these findings may greatly increase the efficiency of SETI searches.  \citet[][]{sheikh2019choosing} describe a linear relationship between the time taken to conduct a SETI search and the maximum drift rate of the search. Our new thresholds built to encompass \textit{most} of the drift rates produced by stable-frequency transmitters on exoplanets can improve the computing costs and times of future searches, such as the one Breakthrough Listen intends to conduct on MeerKAT, as outlined by \citet{Czech_2021}, by nearly three orders of magnitude.

\section*{Acknowledgements}
S.Z.S. acknowledges that this material is based upon work supported by the National Science Foundation MPS-Ascend Postdoctoral Research Fellowship under Grant No. 2138147. 
C.G. acknowledges the support of the Pennsylvania State University, the Penn State Eberly College of Science and Department of Astronomy \& Astrophysics, the Center for Exoplanets and Habitable Worlds and the Center for Astrostatistics.
The Breakthrough Prize Foundation funds the Breakthrough Initiatives which manages Breakthrough Listen.
M.G.L. was funded as a participant in the Berkeley SETI Research Center Research Experience for Undergraduates Site, supported by the National Science Foundation under Grant No.~1950897 and by the NASA Exoplanets Research Program grant 80NSSC21K0575 to UCLA.

\software{\texttt{pandas 1.2.4} \citep[][]{mckinney-proc-scipy-2010, reback2020pandas}, \texttt{astropy 4.2.1} \citep[][]{astropy:2013, astropy:2018}, \texttt{PyAstronomy 0.16.0} \citep[][]{pya}}

\bibliography{references}

\begin{figure*}[ht]
    \centering
    \includegraphics[scale = 0.6]{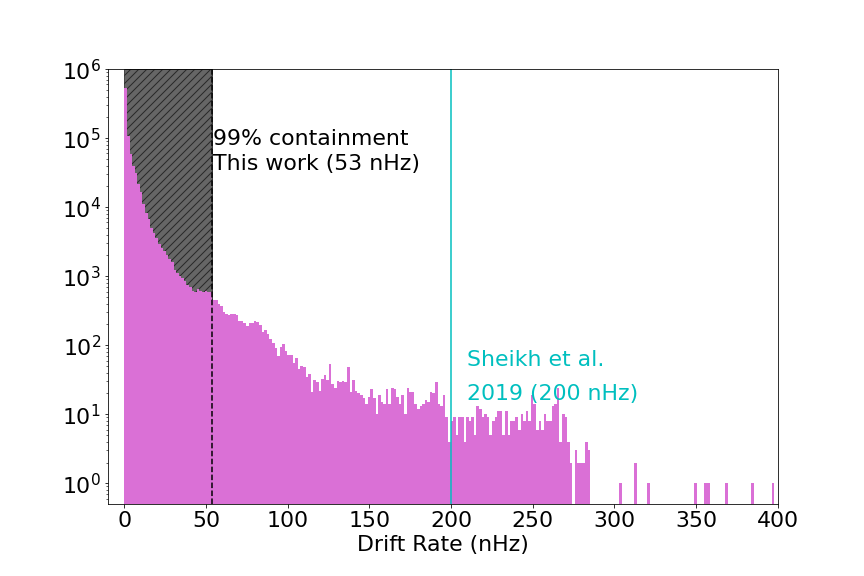} 
    \caption{A histogram of all of the calculated drift rate magnitudes for the 5286 planet sample from the NEA with all inclinations set to $90 ^\circ$. 200 drift rates are calculated for each of the 5286 exoplanets for a total of 1,057,200 drift rates. The dotted line at 53 nHz designate the 99\% boundary: 99\% of calculated drift rates fall between these values. The solid line demonstrates the suggested drift rate upper limits from \citet{sheikh2019choosing}.}
    \label{fig:drift_rate_histogram_2}
\end{figure*}

\appendix 
\label{appendix}
To visualize how observational biases in inclination, period, semimajor axis, and eccentricity affect the distribution of drift rates of a population of exoplanets, we simulated populations of exoplanets with combinations of observed and simulated parameters. As mentioned in section \ref{ssec:comb_sim_with_90}, SETI target selection is also sometimes biased towards known, often transiting, exoplanets. Figure~\ref{fig:drift_rate_histogram_2} shows the distribution of drift rates for the 5286 NEA exoplanets where every inclination is set to $90^\circ$ to mimic a transiting exoplanet population.

\begin{figure*}[ht]
    \centering
    \includegraphics[scale = 0.6]{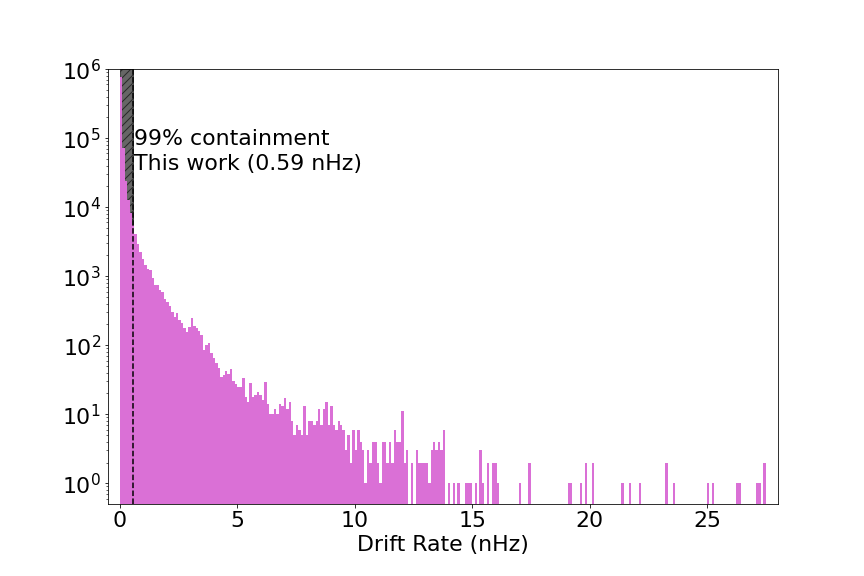} 
    \caption{A histogram of all of the calculated drift rate magnitudes for the 4538 planet sample from the NEA with simulated periods and semimajor axes. 200 drift rates are calculated for each of the 4538 exoplanets for a total of 907,600 drift rates. The dotted line at 0.59 nHz designate the 99\% boundary: 99\% of calculated drift rates fall between these values.}
    \label{fig:drift_rate_histogram_5}
\end{figure*}

Figure \ref{fig:distribution_plots} shows a stark difference between the periods of NEA exoplanets and the simulated population. To see more clearly how our simulated periods affected the distribution of drift rates, we simulated a population using the simulated period distribution from Section \ref{sec:fully_simulated} and solved for corresponding semimajor axes for any NEA exoplanet with a recorded stellar mass, where the rest of the NEA parameters were kept the same. This left a population of 4538 exoplanets, whose drift rates are shown in Figure \ref{fig:drift_rate_histogram_5}. This figure demonstrates that distributions for semimajor axis and period are the most responsible for our simulated results in Section \ref{sec:fully_simulated} moving our 99\% threshold by an order of magnitude.

\begin{figure*}[ht]
    \centering
    \includegraphics[scale = 0.6]{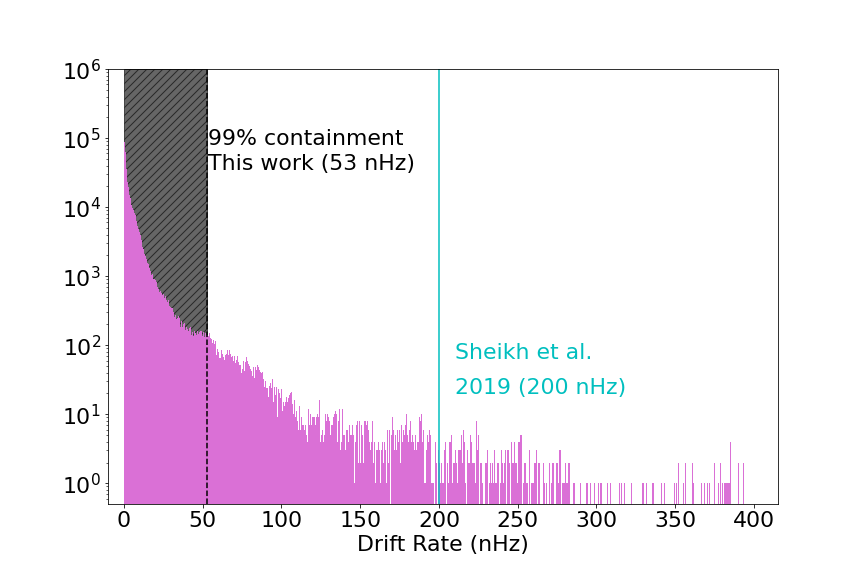}
    \caption{A histogram of all of the calculated drift rate magnitudes for the 5286 planet sample from the NEA with eccentricities sampled from the low eccentricity model. 200 drift rates are calculated for each of the 4286 exoplanets for a total of 1,057,200 drift rates. The dotted line at 53 nHz designate the 99\% boundary: 99\% of calculated drift rates fall between these values. The solid line demonstrates the suggested drift rate upper limits from \citet{sheikh2019choosing}.}
    \label{fig:drift_rate_histogram_6}
\end{figure*}

Figure \ref{fig:distribution_plots} shows a slight difference between eccentricities recorded in the NEA and the low eccentricity model. To see how this difference affected the drift rate distribution, we simulated a population of exoplanets using the simulated low eccentricity population model for the NEA population where the rest of the NEA parameters were kept the same. The drift rate distribution for these exoplanets is shown in Figure \ref{fig:drift_rate_histogram_6}. 

\begin{figure*}[ht]
    \centering
    \includegraphics[scale = 0.6]{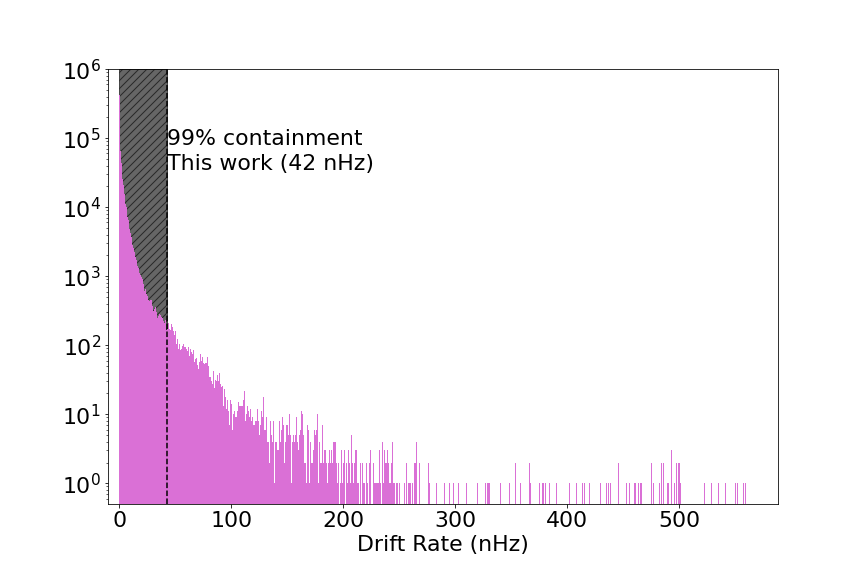} 
    \caption{A histogram of all of the calculated drift rate magnitudes for the 5286 planet sample from the NEA with inclinations sampled uniformly in cosine. 200 drift rates are calculated for each of the 4286 exoplanets for a total of 1,057,200 drift rates. The dotted line at 42 nHz designate the 99\% boundary: 99\% of calculated drift rates fall between these values.}
    \label{fig:drift_rate_histogram_7}
\end{figure*}

As mentioned in Section \ref{ssec:sim_parameters}, observable inclinations are heavily affected by methods of discovery and also differ significantly from our simulated population of inclinations. Thus, we simulated a population of exoplanets with inclinations drawn from a uniform distribution of $ -1 \leq cos(i) \leq 1$ with the rest of the parameters kept the same from the NEA. The effect this difference has on the distribution of drift rates is visualized in Figure \ref{fig:drift_rate_histogram_7}.

\end{document}